\documentclass[twoside,slac_one,lettersize]{revtex4}
\usepackage{graphicx}
\usepackage{fancyhdr}
\usepackage{amsmath} 
\usepackage{bm}
\usepackage{amsxtra}
\usepackage{amssymb}
\usepackage{amsthm}
\usepackage{latexsym}
\usepackage{lscape}
\usepackage{cancel}

\newcommand{\sect}[1]{Sec.~\ref{#1}}

\newcommand{\fig}[1]{Fig.~\ref{#1}}
\newcommand{\tableref}[1]{Table~\ref{#1}}

\begin{document}

\title{Searches For Fourth Generation Quarks With The CMS Detector}

%

\author{Michael M. H. Luk, on behalf of the CMS Collaboration}
\affiliation{Department of Physics and Astronomy, Brown University, Providence, RI, USA}

\begin{abstract}
We summarise the analyses that search for fourth generation quarks at the Central Muon Solenoid (CMS) experiment. Such particles provide a natural extension to the Standard Model (SM) and are still consistent with precision electroweak measurements. Our searches are not limited to fourth generation chiral quarks and are relevant to many Beyond the Standard Model theories. No excess over the expected SM background is observed in any of these analyses and limits are set on the masses of the $b^\prime$ and $t^\prime$ quarks at $95\%$ confidence level at $361$ GeV/$c^2$ and $450$ GeV/$c^2$, respectively.
\end{abstract}
\maketitle 
\thispagestyle{fancy}

\section{Introduction}
Many Beyond the Standard Model theories predict a heavy family of quarks \cite{bsmmodels1,bsmmodels2}. These additional quarks can offer solutions to some outstanding theoretical questions; most notably, they may allow sufficient CP violation to explain the matter-antimatter asymmetry and may also resolve the issue of naturalness of the Higgs mass. Further, the addition of these particles is still consistent with the stringent limits set by precision electroweak experiments \cite{ewkconstraints0, ewkconstraints1,ewkconstraints2,ewkconstraints3}. Lower mass limits on the masses of the $b^\prime$ and $t^\prime$ have been set at $m_{b^\prime} > 385$ GeV/$c^2$ \cite{bprimelimit} and $m_{t^\prime} > 358$ GeV/$c^2$ \cite{tprimelimit1, tprimelimit2} at the $95\%$ CL by CDF and D$0$.

These proceedings cover three analyses that search for strongly pair-produced fourth generation quarks at the CMS detector at $\sqrt{s}=7$ TeV. Specifically, these analyses look for a heavy family of quarks which decay into SM particles and proceed through the following decay channels:
\begin{eqnarray}
b^\prime \bar{b}^\prime &\longrightarrow & tW^- \bar{t} W^+ \\
T^\prime \bar{T}^\prime &\longrightarrow & tZ \bar{t}Z   \\
t^\prime \bar{t}^\prime &\longrightarrow & bW^+ \bar{b}W^-
\end{eqnarray}
Each analysis and their results shall be discussed in the following. 

\subsection*{CMS Detector}
The fourth generation searches are carried out at the Large Hadron Collidor (LHC) using the data recorded by the CMS detector at $7$ TeV centre of mass energy. The CMS detector is a large-solid-angle magnetic spectrometer with a superconducting solenoid that applies an axial magnetic field of $3.8$T. The tracks from charged particles are recorded in the tracking detector, which consists of silicon pixel and silicon strip detectors. The silicon tracker covers the azimuthal range between $0$ and $2\pi$ and the pseudorapidity range $|\eta|\leq 2.5$ (where $\eta \equiv - \ln [\tan (\theta/2)]$ with $\theta$ the polar angle relative to the anticlockwise proton beam direction). The transverse and longitudinal directions are defined as the directions perpendicular and parallel to the beam axis, respectively. 

The tracker is enclosed by a lead tungstate crystal electromagnetic calorimeter (ECAL), with a lead-silicon preshower detector in the end-caps. In turn this is surrounded by a brass-scintillator hadronic calorimeter (HCAL). Muons are identified by the gas-ionisation muon chambers that are embedded in the steal return yoke outside the solenoid. Together these components make up a near-hermetic detector which can provide high resolution measurements. A two-tier hardware/software trigger system is then implemented to select interesting proton-proton collision events. A more detailed description of the CMS detector can be found elsewhere \cite{cmsdet}.

\section{Search for a Heavy Bottom-like Quark}
We present the search for a pair-produced heavy bottom-like quark, $b^\prime$, using the data collected in $2010$ at the CMS detector, amounting to $34$ pb$^{-1}$ of integrated luminosity \cite{bprimenote}. The $b^\prime$ quark is assumed to decay into a top, $t$, and $W$ boson. The $pp\rightarrow b^\prime \bar{b^\prime} \rightarrow  tW \bar{t}W$ process can be identified by the distinctive signature in the trilepton and same-sign dilepton channels, with the leptons coming from decaying $W$-bosons.


\subsection{Event Selection}
Events are required to have either two same-sign leptons or three leptons (two of which are required to be oppositely charged). Events containing two same-sign dilepton (trilepton) channel are required to have four (two) or more jets. A veto is placed on events with a same-flavour, oppositely-charged lepton pair (either $e$ or $\mu$) that have invariant mass near the $Z$ boson mass ($|M_{l^+l^-} - M_Z|<10$ GeV/$c^2$). This suppresses the Z boson background. Additionally, since the significant electron charge misidentification for electrons, same-sign electron pairs whose invariant mass is within $10$ GeV/$c^2$ of the $Z$ boson mass ($|M_{e^\pm e^\pm} - M_Z| < 10$ GeV/$c^2$) are also discarded. 

Finally the cut $S_T > 350$ GeV/$c$ is imposed where $S_T \equiv \sum p_T(\text{jets}) + \sum p_T(\text{leptons}) + \cancel{E}_T$, with $\cancel{E}_T$ the missing transverse energy.

\begin{figure}[h]
\centering
\includegraphics[width=0.68\columnwidth]{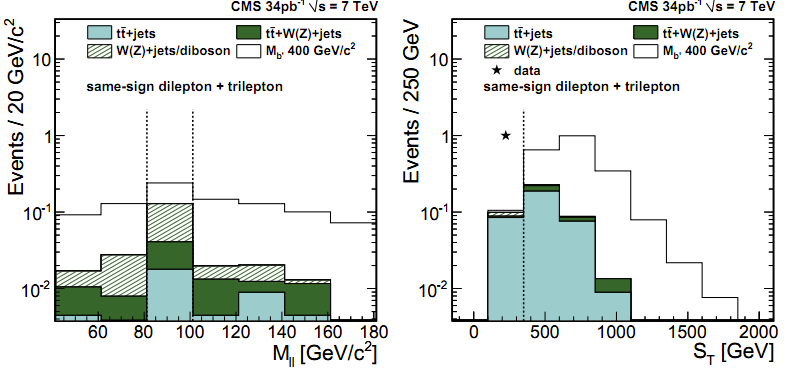}
\caption{Left: Shows the invariant mass distribution of two opposite-charge muons or any-charge electrons with all selection requirements imposed except $|M_{l^+l^-} - M_Z|<10$ GeV/$c^2$ (between the two dotted lines). Right: Shows the $S_T$ distribution of the same-sign dilepton and trilepton channels. All selection requirements except $S_T>300$ GeV/$c$ (dotted line) are imposed.}
\label{figure:bprimeplots}
\end{figure}

The distributions of the invariant mass of the leptons and the $S_T$ variable with their respective cuts are shown in \fig{figure:bprimeplots}. The shaded contributions in the plots correspond to the expected SM background, whilst a hypothetical contribution from a $b^\prime$ quark of mass $400$ GeV/$c^2$ is also shown (unshaded) in each of the figures.  All selection requirements are applied in these plots except for their respective cuts  (vertical dotted line).

\subsection{Event Yield}
The event yield and selection efficiencies, $\epsilon=n_s / (\sigma \mathcal{L})$, where $n_s$ and $\sigma$ are the expected number of events and the cross section for the signal process respectively, at $34$ pb$^{-1}$ can be found in \tableref{table:bpeventyield}. The background yield, including both same-sign dilepton and trilepton channels, is estimated by a data driven method to be $0.32$ events in the signal region (full details in \cite{bprimenote}). This is in good agreement with a direct counting of simulated background events, $0.33$ events, that satisfy the selection criteria. The small difference between these two yields is included in the systematic uncertainties. 

The background yield is taken to be $0.32$ events in the signal region and has a total relative uncertainty of $65\%$. No events were observed in the data, consistent with the background only hypothesis. Relaxing the $S_T$ cut to $200$ GeV/$c$ and jet multiplicity to $1$, introduces 2 events, which is again consistent with the corresponding background yield of $0.69$ events. 

\begin{table}[h]
\centering
\includegraphics[width=0.65\columnwidth]{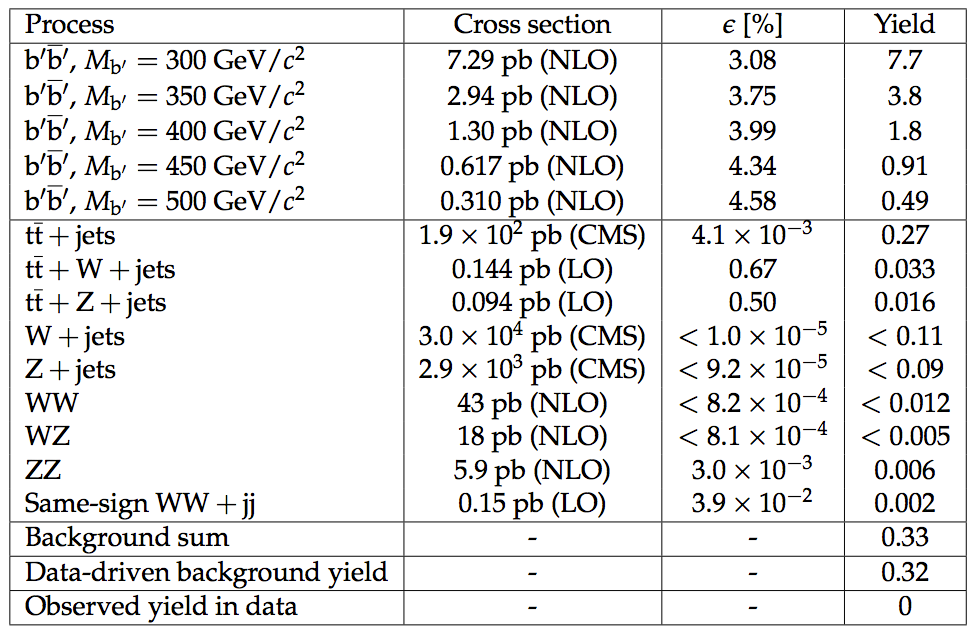}
\caption{Shows the expected event yields and their selection efficiencies for hypothetical $b^\prime$ signal, SM background, and observed in data.}
\label{table:bpeventyield}
\end{table}

\subsection{Results and Limit Setting}
\begin{figure}[h]
\centering
\includegraphics[width=0.68\columnwidth]{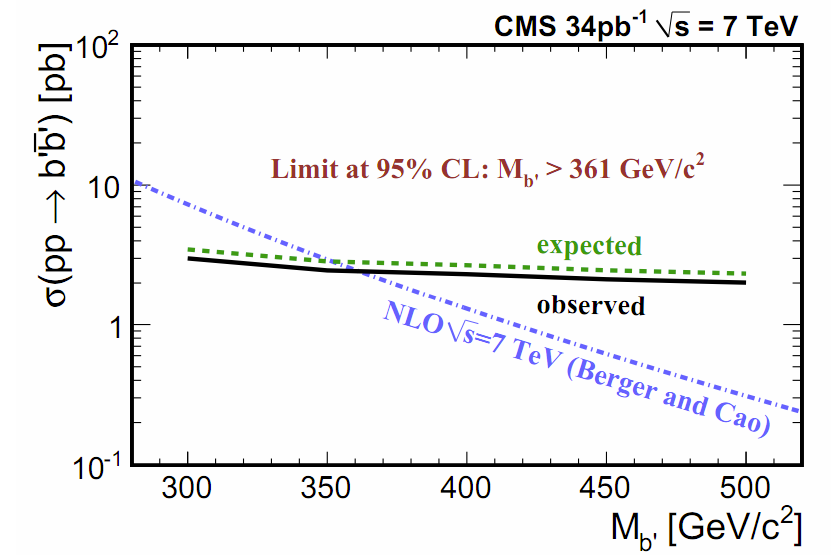}
\caption{Shows the exclusion limits at $95\%$ CL on the cross section of $b^\prime$ pair production. The NLO theoretical production cross sections at $7$TeV (dotted blue) crosses the observed limit (solid black) at $361$ GeV/$c^2$ setting a $95\%$ exclusion mass limit at this value. The expected limit (dotted green) represents the limit for a background only hypothesis with the available sample size.}
\label{figure:bprime_ex}
\end{figure}

From the estimated background and signal yields, a limit on the cross section is set at $95\%$ CL for each mass hypothesis. A Bayesian method is implemented for the limit computation with log-normal priors for the nuisance parameters \cite{bprimelimcalc}. The exclusion plot at $34$ pb$^{-1}$ is shown in \fig{figure:bprime_ex} and excludes a $b^\prime$ mass less than $361$~GeV$/c^2$ at $95\%$ CL. 

The search for a pair-produced heavy bottom-like quark in proton-proton collisions  at the CMS detector at $\sqrt{s} = 7$ TeV is presented. The process studied is the low yield $pp\rightarrow b^\prime \bar{b^\prime} \rightarrow  tW \bar{t}W$. No excess over the SM was found in the $34$ pb$^{-1}$ of data collected during 2010. A $b^\prime$ quark mass range from $255$ to $361$ GeV/$c^2$ is excluded at $95\%$ CL.

\section{Search for a Top-like Quark Decaying to a Top Quark and a Z boson}
\label{section:tprimetz}
We present the search for a pair-produced heavy top-like quark, $T$, using $191$ pb$^{-1}$ of integrated luminosity collected between 2010 and 2011 \cite{tprimetznote}. The $T$ quark is assumed to decay into the SM top quark and $Z$ boson. The flavour changing neutral current decay $pp\rightarrow T \bar{T} \rightarrow  tZ \bar{t}Z$ is considered by selecting events with two leptons from a Z decay and an additional isolated, charged lepton. 

Within the SM, flavour changing neutral currents are highly suppressed but there are many beyond the SM theories which allow a vector-like quarks to couple at tree-level to FCNCs. Such models can allow the $T\rightarrow tZ$ decay to be the dominant channel \cite{fcncvecquarks1,fcncvecquarks2,fcncvecquarks3}. 

The events considered follow the decay chain $pp\rightarrow T \bar{T} \rightarrow  tZ \bar{t}Z \rightarrow b W^+ Z \bar{b}  W^- Z Z$. This has a very clean signal in the trilepton channel where the expected background is small - with two leptons coming from the $Z$ decay, and one from the leptonic $W$ decay.

\subsection{Event Selection}
Events are required to have at least one well-reconstruction interaction vertex, a leptonically decaying $Z$ boson, an additional isolated lepton and at least $2$ jets. A $Z$ boson that decays into leptons is identified by two same-flavoured oppositely-charged leptons with invariant mass $60 < M_Z < 120$ GeV$/c^2$. Further, the third isolated lepton is required in the event to identify a leptonically decaying $W$ boson.

Finally, a cut on the residual $S_T\equiv \sum_{i\neq 1,2} p_T (\text{jets}) + \sum_{i \neq 1,2} p_T (\text{leptons})$ of greater than $80$ GeV$/c$ cut is used to reduce the SM backgrounds. 

\begin{figure}[h]
\includegraphics[width=0.35\columnwidth]{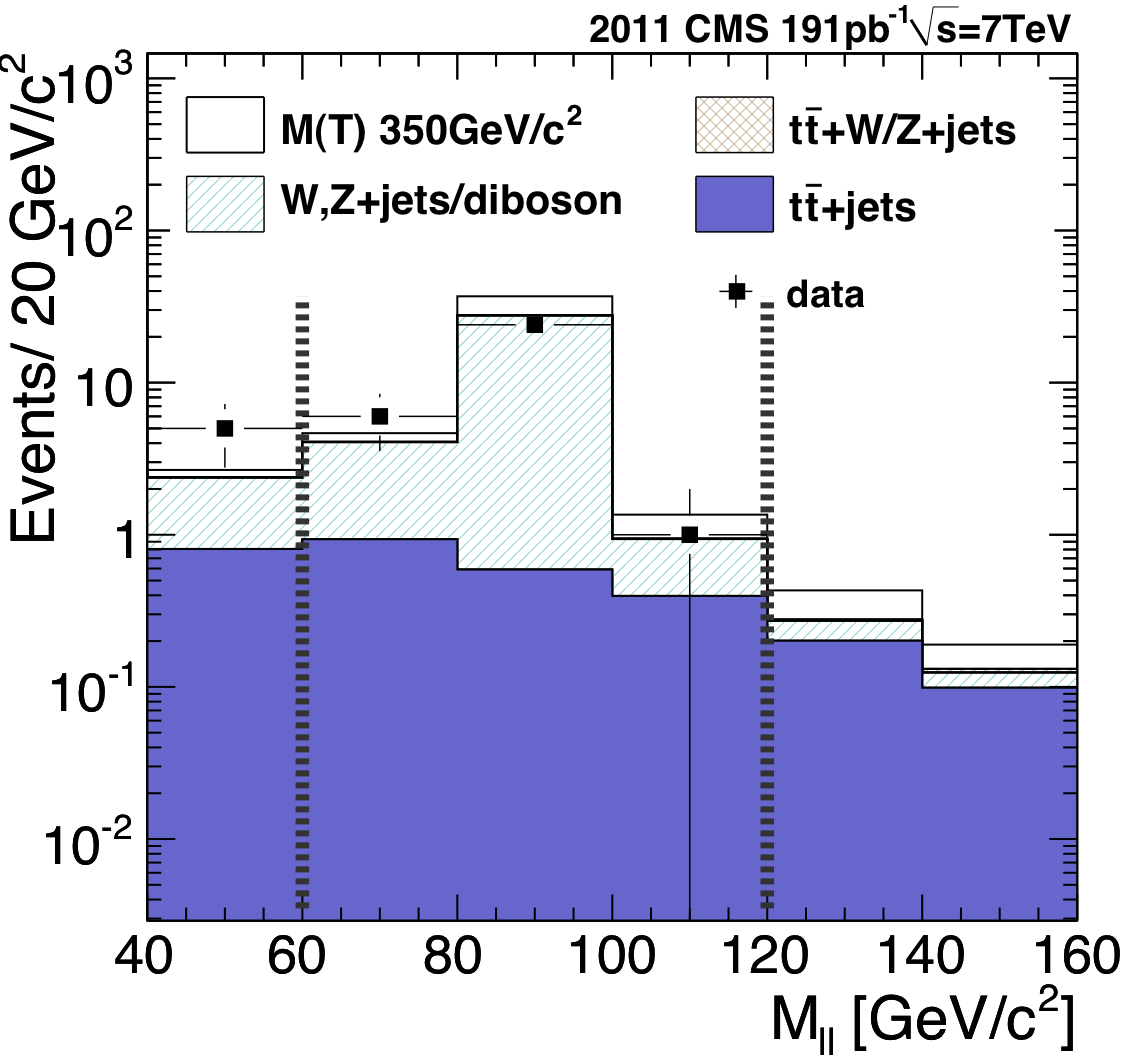} 
\includegraphics[width=0.35\columnwidth]{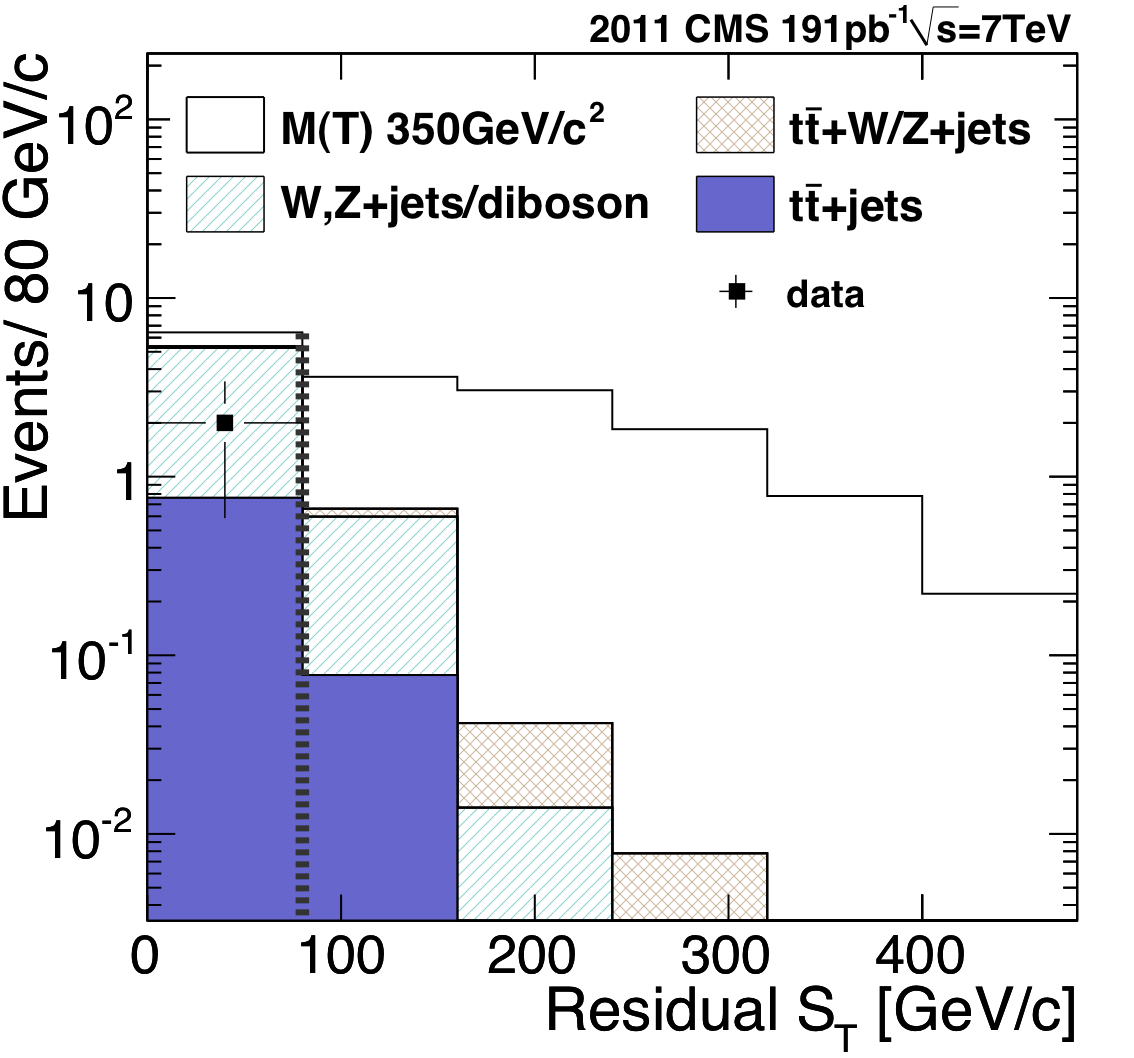} 
\caption{Left: Shows the invariant mass of the two like-flavoured oppositely-charged leptons. The range between the two vertical dotted lines represents the signal region. Right: Shows the residual $S_T$ distribution. All selection except the cut on the respective variable shown in each plot is applied. } 
\label{figure:Tplots}
\end{figure}

The invariant mass of the leptons and the $S_T$ variable along with their respective cuts are shown in \fig{figure:Tplots}. A hypothetical contribution from a $T$ quark of mass $350$ GeV/$c^2$ is also shown (unshaded) in each of the figures. The shaded portion corresponds to the expected SM background. All selection requirements are applied in these plots, except the corresponding threshold (dotted lines) of their respective plots. 

\subsection{Event Yield}
The event yield and signal efficiencies  at $191$ pb$^{-1}$ can be found in \tableref{table:Teventyield}. The event yields for the various samples are calculated from a combination of simulation estimates and data measurements. The estimated background yield in the signal region is $0.73$ events with a total uncertainty of $\pm 0.31$ events (a full description of the systematics can be found in \cite{tprimetznote}). No events that passed all selection criteria were observed in the data - this is consistent with a SM background only hypothesis.

\begin{table}[h]
\includegraphics[width=0.68\columnwidth]{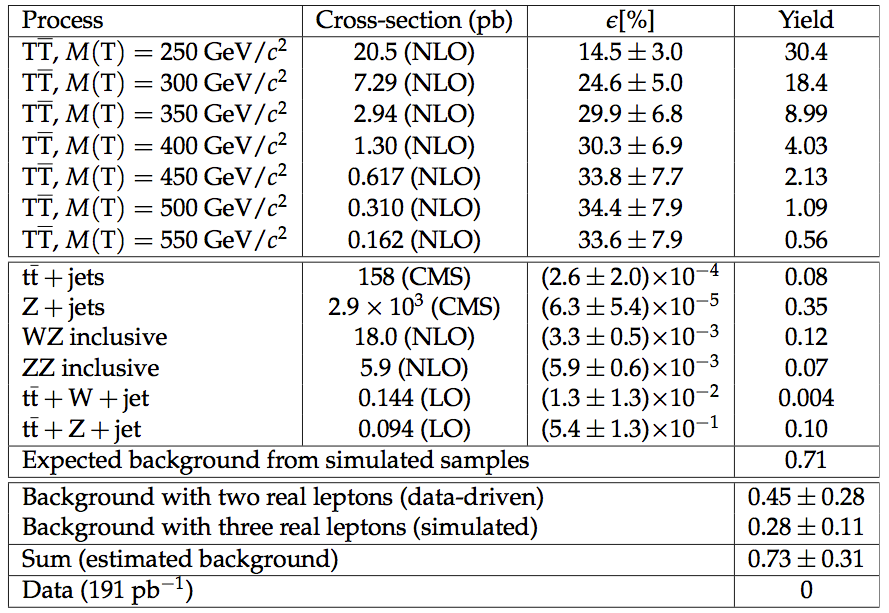}
\caption{Shows the event yields normalised to $191$ pb$^{-1}$ of the various SM backgrounds, the expected signal yields at various mass points of a hypothetical $T$ quark and the observed number of events in the data.}
\label{table:Teventyield}
\end{table}

\subsection{Results and Limit Setting}
\begin{figure}[h]
\includegraphics[width=0.68\columnwidth]{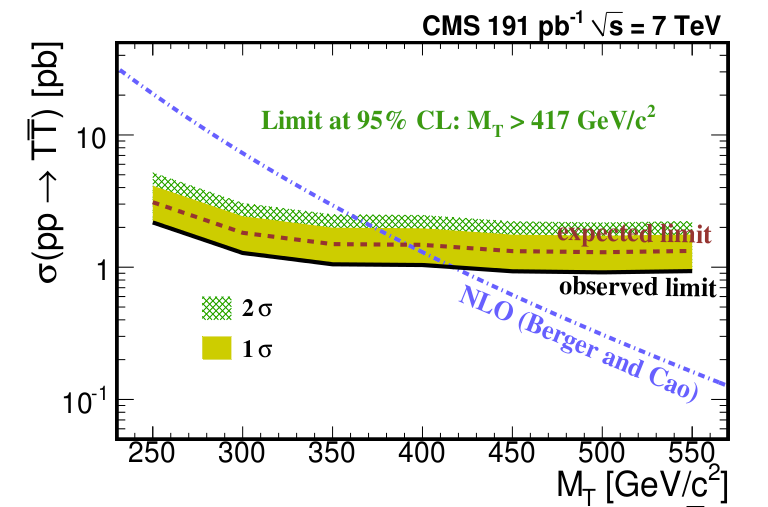}
\caption{Shows the $95\%$ CL limits set on a pair-produced heavy top-like quark using $191$ pb$^{-1}$ of integrated luminosity. The observed limit is shown in black and the dotted red line is the expected limit for a background only hypothesis. The NLO theoretical cross section at centre of mass energy $\sqrt{s}=7$ TeV is shown in dotted blue.} 
\label{figure:Tlimits}
\end{figure}

From the estimated background and signal yields \tableref{table:Teventyield}, a limit on the $T \bar{T}$ cross section is set at $95\%$ CL for each mass hypothesis. A Bayesian method is used to calculate the upper cross section limit for each $T$ mass considered - this result is shown in \fig{figure:Tlimits}. The number of observed events in data is consistent with the SM background processes only and a lower bound on the $T$ quark mass is set at $417$ GeV$/c^2$, assuming a $100\%$ branching ratio in the $T\rightarrow tZ$ channel. 

The search for a pair-produced heavy top-like quark in proton-proton collisions at the CMS detector is presented. Using $191$ pb$^{-1}$ of data collected at centre-of-mass energy $7$ TeV no excess over the expected SM processes was found. We exclude a strongly pair-produced $T$ quark, assuming a $100\%$ branching ratio in the $T\rightarrow tZ$ channel, with mass less than $417$ GeV/$c^2$ at $95\%$ CL.

\section{Search for a Fourth-Generation $t^\prime$ Quark in the Lepton-plus-Jets Channel}

We present the search for a pair-produced heavy top-like quark, $t^\prime$ decaying in the a lepton plus jets channel in proton-proton collisions at $7$ TeV in the CMS detector \cite{tprimebwnote}. 

To agree with precision electroweak measurements, the mass splitting of such a fourth generation top-like quark and its bottom partner must be small ($m_{t^\prime} - m_{b^\prime} < 50$ GeV/$c^2$  \cite{ewkconstraints1,ewkconstraints2,ewkconstraints3}); thus, the decay to a $b^\prime$ quark plus $W$ boson is heavily disfavoured. For this analysis, we assume that the $t^\prime$ is pair-produced and decays predominantly in the $t^\prime\rightarrow bW$ channel. Recent direct searches have put the lower limit of such a $t^\prime$ quark mass to be above $358$ GeV/$c^2$ \cite{tprimelimit1,tprimelimit2}. 


Assuming strongly produced $t^\prime$ quarks that decay dominantly in the $t^\prime\rightarrow bW$ channel, a $450$ Gev/$c^2$ lower limit on the $t^\prime$ mass is set at $95\%$ CL. 

The full decay chain of events considered is the $pp\rightarrow t^\prime \bar{t^\prime} \rightarrow  bW^+ \bar{b}W^- \rightarrow b l\nu  \bar{b} q \bar{q}$. Events with one electron, $e+$jets, and one muon, $\mu+$jets, are analysed separately, with the event selection optimised in each channel to maximise signal sensitivity. Statistical results from both channels are then  combined to give a final limit.

There are several SM process that give rise to a lepton-plus-jets signature. Furthermore, the SM $t\bar{t}$ production process is an irreducible background whose final state particles exactly match that of the signal. 

Therefore, two discriminating variables, the reconstructed $t^\prime$ mass, $m_{fit}$ and the scalar sum of transverse energies of the lepton, jets and missing $p_T$, $H_T$, are used to distinguish the $t\bar{t}$ production and a potential pair-produced fourth generation $t^\prime$ quark signal.

\subsection{Event Selection}

All are required to have at least $4$ jets with $p_T > 120,90,35,35$ GeV/$c$ and $|\eta|<2.4$, missing $p_T > 20$ GeV/$c$ and more than one b-tagged jet (i.e. a jet that is determined to be a b-jet). Additionally, both channels require events to have a primary vertex with $|z|< 24\text{cm},|r|<2\text{cm}$, where $z$ ($r$) is the distance from the centre of the detector parallel (perpendicular) to the beam line. In the $\mu+$jets channel, the muon is required to have $p_T> 35$ GeV/$c$, $|\eta|<2.1$. In the $e+$jets channel, the electron is required to have $p_T > 30,35,45$ GeV/$c$ (to match the trigger thresholds that were updated during data collection) with $|\eta|<2.5$ excluding a transition region in the detector at $1.44< |\eta| <1.56$. In both cases the lepton is required to have a transverse impact parameter $|d_{xy}|<0.02 \text{cm}$ and a longitudinal impact parameter $|d_z| < 1 \text{cm}$. Full details of the event selection can be found in \cite{tprimebwnote}.

\subsection{Event Yield}
\label{section:tprimesel}
The event yields and signal selection efficiencies, which include the branching ratio, can be found in \tableref{table:tprimeeventyield}. The uncertainties considered include the uncertainty in the jet energy scale, the lepton selection efficiency, the uncertainty in the b-tag efficiency, and the statistical uncertainty from the Monte Carlo generation. In the limit setting procedure uncertainties in the cross sections are also included. Full description of the uncertainties and how they are handled can be found in \cite{tprimebwnote}.

\begin{table}[h]
\centering
\includegraphics[width=0.5\columnwidth]{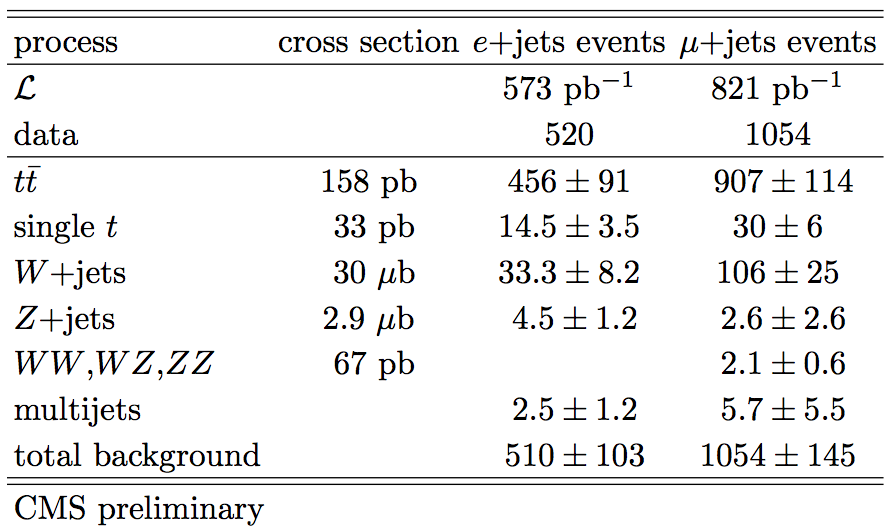}
\includegraphics[width=0.5\columnwidth]{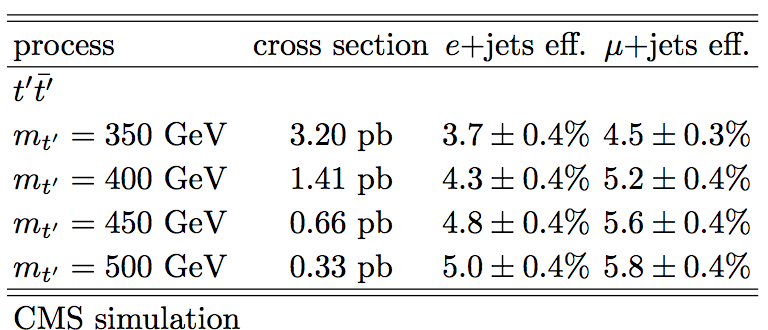}
\caption{Shows the selection efficiencies and event yields with integrated luminosity $573$ pb$^{-1}$ \& $821$ pb$^{-1}$ in the $e+$jets and $\mu+$jets channels respectively. The $t\bar{t}$ cross section comes from CMS measurement \cite{ttbarxs}, and $t^\prime\bar{t^\prime}$ cross section is calculated by HATHOR \cite{hathor}.}
\label{table:tprimeeventyield}
\end{table}

\subsection{Mass Reconstruction}
Since the SM $t\bar{t}$ background is irreducible the reconstructed mass is used as a discriminating variable. The mass reconstruction proceeds with a kinematic fitter, which is implemented to fully reconstruct the event. Imposing the three constraints that $M_W = m(l\nu)$, $M_W = m(qq)$ and $m(l \nu b) = m(qqb)$, we can use the measurements of the final state particles to reconstruct the event.

The reconstructed objects in the event are the charged lepton, the missing transverse momentum, and the four or more jets. The neutrino's transverse momentum is reconstructed using momentum conservation and its longitudinal momentum can be deduced up to a two-fold ambiguity by imposing the above constraints.

Taking only the four leading (highest $p_T$) jets, there are $12$ combinations of matching the reconstructed jets to partons (24 combinations including the ambiguity in the longitudinal momentum solution).  From this set of combinations, the one which returns the smallest $\chi^2$ is chosen, with a $\chi^2$ value for each combination determined by the goodness of fit to the constraints.

\begin{figure}[h]
\includegraphics[width=0.68\linewidth]{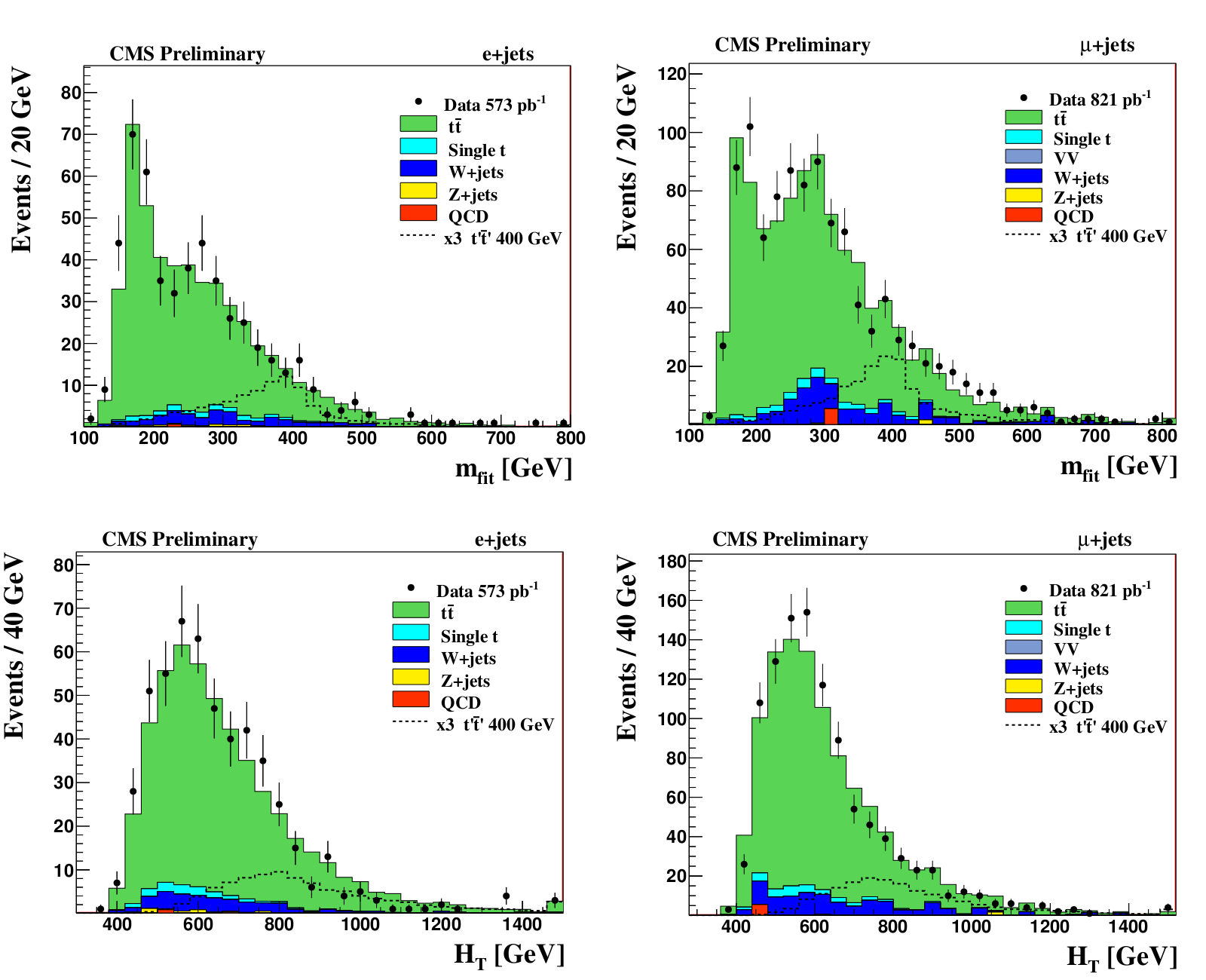}
\caption{Left: Shows the $e+$jets fitted mass and $H_T$ variables at $573$~pb$^{-1}$. Right: Shows the $\mu+$jets fitted mass and $H_T$ variables at $821$~pb$^{-1}$. Plots include all selection mentioned in  \sect{section:tprimesel}.}
\label{figure:tprimefits}
\end{figure}

In principle, considering more jets increases the likelihood that the correct combination is included in the set considered at the cost of introducing more combinations to pick from. A study was done for this analysis, which showed that $5$~leading jets was the best trade off between these two factors. As such, the $e+$jets channel chooses the lowest $\chi^2$ combination out of the set of all combinations made from the $5$ leading jets. In the $\mu+$jets channel, the $4$~leading jets are considered by default, except in the case when the $5^{th}$ jet is the btagged jet, in which case it replaces the $4^{th}$ jet. This difference in jet-parton assignment strategy in the two channels accounts for the difference in shape of the fitted mass distribution \fig{figure:tprimefits}. The secondary hump in the $\mu+$jets fitted mass is due to a leading jet that is used in the fit coming from high momentum initial or final state radiation rather than from a $t\bar{t}$ decay.
  
\subsection{Results and Limit Setting}

For each sample, a $2D$ plot is made of $m_{fit}~vs~H_T$, see \cite{tprimebwnote}. The number of events in the signal, data and background histograms are then compared bin-by-bin by a statistical package \cite{tpstats1,tpstats2,tpstats3}. The CLs method is used to calculate lower limits on the cross section at each $t^\prime$ quark mass considered. In the $e+$jets channel, using $573$ pb$^{-1}$ of data, an observed lower limit on the $t^\prime$ mass is set at $431$ GeV/$c^2$ at $95\% CL$. For the $\mu+$jets channel, using $821$ pb$^{-1}$ of data gives an observed limit of  $m_{t^\prime} > 422$ GeV/$c^2$ at $95\% CL$.

\begin{figure}[h]
\includegraphics[width=0.68\columnwidth]{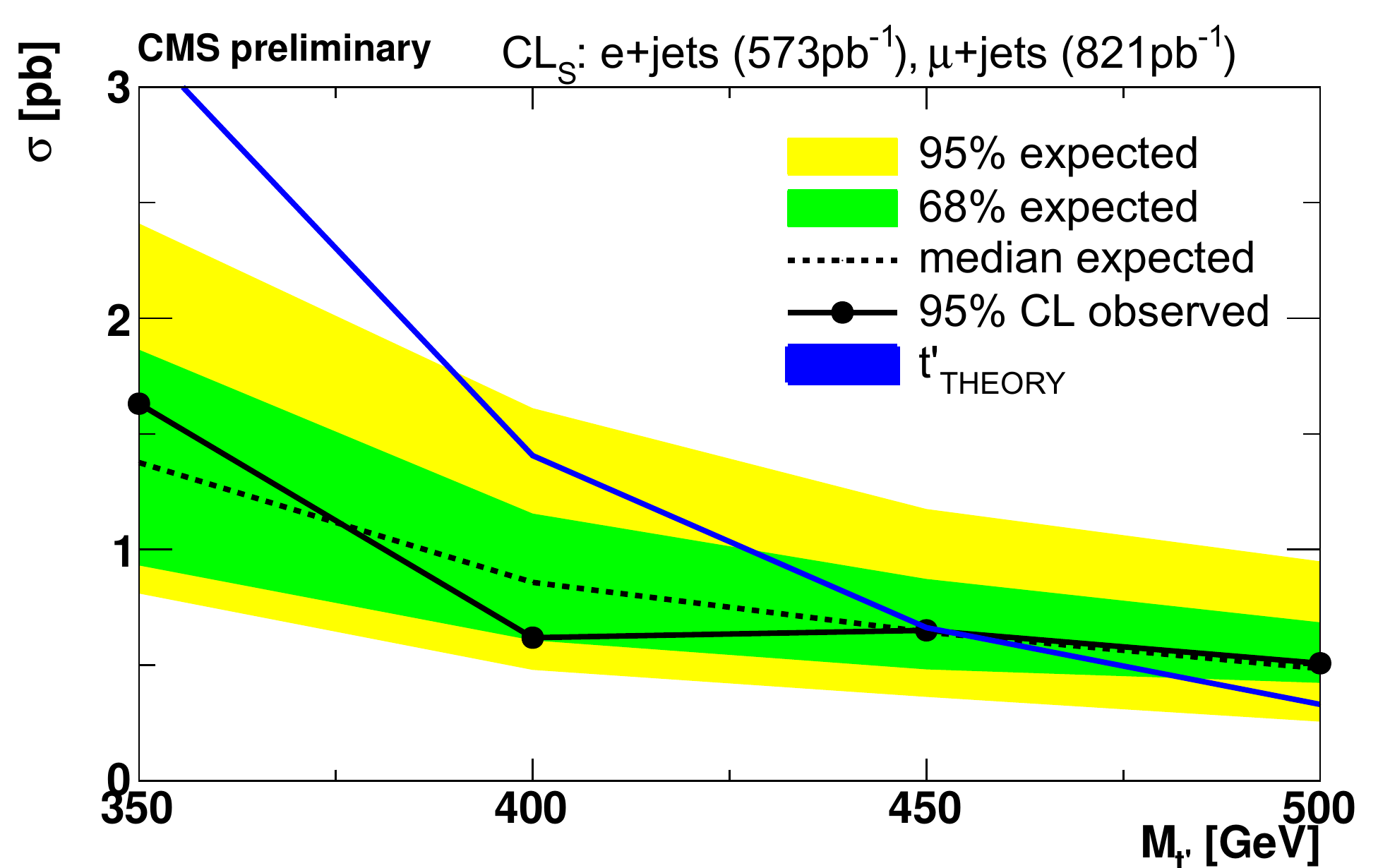}
\caption{Shows the combined lepton-plus-jets cross section limit as a function of mass for the pair-produced $t^\prime$ quark at $\sqrt{s} = 7$ TeV. We report an observed limit of $m_{t^\prime} > 450$ GeV$/c^2$ at $95\% CL$.}
\label{figure:tprimelimit}
\end{figure}

The two lepton channels are combined and lower cross section limit is again set at each mass point, \fig{figure:tprimelimit}. A strongly produce $t^\prime$ quark with $m_{t^\prime}$ is excluded up to $450$ GeV/${c^2}$ in the combined lepton plus jets channel at $95\%$ CL.




In conclusion, the search for a pair-produced heavy top-like quark in proton-proton collisions at the CMS detector is presented. Using $573$ pb$^{-1}$ of data in the $e+$jets channel and $821$ pb$^{-1}$ of data in the $\mu+$jets channel collected at $\sqrt{s}=7$ TeV no excess over the expected SM processes was found. We exclude a pair-produced $t^\prime$ quark, with $100\%$ branching ratio in the $t^\prime\rightarrow bW$ channel, with mass less than $450$ GeV/$c^2$ at $95\%$ CL.

\section{Final Comments}
No evidence for fourth generation quarks is observed at the CMS detector at $7$TeV. A pair-produced heavy bottom-like quark, $b^\prime$ has been excluded at to $95\%$ CL for masses between $255$ to $361$ GeV/$c^2$ with the $34$ pb$^{-1}$ of integrated luminosity recorded in 2010.   A lower mass limit on the $T$ quark which decays into a SM top quark and a $Z$ boson has been set at $95\%$ CL and excludes strongly pair-produced $T$ quarks up to $417$GeV/$c^2$ with $191$ pb$^{-1}$ of data. Finally, a limit on a strongly pair-produced $t^\prime$ quark that decays into the a $b$ quark and a $W$ boson has been excluded up to $450$ GeV/$c^2$ at $95\%$ CL using $573$ pb$^{-1}$ and $821$ pb$^{-1}$ of integrated luminosity in the $e+$jets and $\mu+$jets channels, respectively.

\end{document}